# Energetic Controls are Essential

*'New and Notable Note'*
*concerning a paper by Davis et al, (1)*

Robert S Eisenberg

February 1, 2020

Department of Applied Mathematics
Illinois Institute of Technology

Department of Physiology and Biophysics
Rush University Medical Center

Chicago IL
USA

Methodological controls seem an unlikely subject for a 'New and Notable' note but Davis, et al, (1) are an exception. Davis, et al, investigate a promising new method for non-invasive control of calcium currents in individual cells in the nervous system by the selective heating of nanoparticles (2, 3) and show that simple physical laws, properly applied, explain what is happening, and so can be a foundation for constructing improved methods and techniques.

Davis, et al, investigate the radiofrequency heating of superparamagnetic iron oxide nanoparticles (SPIONs). The heating of SPIONs has been explained in various ways, often with a heat confinement mechanism. Davis, et al, evaluate a simple explanation (in essence Fourier's heat law without extra heat confinement) and find no significant difference between the surface temperature of the nanoparticle and the temperature of the surrounding fluid when radio frequency alternating magnetic fields are applied. They use custom built flurometers, measuring two different dyes, analyzing results to show that no special surface properties are needed to understand the results. Classical analysis of heat flow is enough. Measurements of surface temperature with optical thermometry are examined and politely reinterpreted.

As so often in classical biophysics, doing the right experiments—with carefully designed custom instrumentation, and controlled experimentation that allow quantitative reproducible results—shows how well established physical principles govern the results of our biophysical experiments.

Davis, et al, (1) show just how important quantitative measurement is in the understanding of experimental and biological phenomena. Engineering does not exist without numbers and it seems likely that numbers are just as important in biophysics as in engineering.

Modern molecular and structural biology have been so remarkably productive that we sometimes forget that both are qualitative sciences that can inadvertently substitute description for quantitative understanding. This forgetfulness is ironic. Structural biology, for example, depends for its power on magnificent quantitative sciences: physics, engineering and mathematics. It is easy to forget the quantitative science required to produce x-rays (for structural analysis of proteins) from synchrotrons that operate at some 7 billion volts, that move electrons at 99.99999999% of the speed of light, with currents of some 200 milliamps and that put electrons and x-rays where they are supposed to be. Structural biology could not exist without the tools built for them by physicists and engineers …. and mathematicians. It is very easy for structural biologists to ignore the applied mathematics that has taken x-ray crystallography from an art to a routine science. Fourier transforms are not even in the vocabulary of many younger biophysicists. That is as it should be: a sign of successful engineering is that the user is unaware of its presence.

As successful as molecular and structural biology have been, engineering has been more so. Semiconductor engineering has been an astounding success story because of quantitative science and mathematics. Never in human history has a technology improved by a factor of roughly one billion, in some fifty years or so.(4-6) The smartphones most of us carry contain some $10^{12}$ transistors that switch in some $10^{-9}$ sec with zero error rates in the crucial parts of their arithmetic logic units. Nothing like this was available even thirty years ago. Nothing like this was foreseen by scientists 50 years ago, although Captain Kirk in StarTrek pretended otherwise, with his communicators and tricorders, as imagined by its creator Gene Roddenberry.

What is so remarkable is that our smartphones are not pretend. They work reliably because they depend on physics and mathematics that can be used for design.(7) Design is possible for circuits with $10^{12}$ components, and many more connections, only because the physics and mathematics describing integrated circuits is nearly exact. The electrodynamics part of the description is exact (8), even within atoms, and its expression as Kirchhoff's laws is nearly exact.(9) Kirchhoff's law requires only a few adjustable (structural) parameters ('stray capacitances') to be an accurate robust description.(10)

The success of semiconductor physics depends on the understanding that atoms follow well known physical laws in which electrodynamics, friction and thermal motion dominate.(11, 12) What a remarkable contrast there is between the wonderful qualitative success of molecular and structural biology and the essentially quantitative success of semiconductor electronics and its computational support.

It is tempting for biologists to treat ions and protein sidechains as friends, moving according to their wishes. But atoms do not move that way. They move according to the laws of physics, in thermal, nearly Brownian motion, with atoms moving at (more or less) the speed of sound, some 1,300 meters per second in water, or roughly three or four water diameters (really 10 Angstroms) every $10^{-12}$ sec, moving in condensed phases with almost no empty space. The atoms of condensed phases experience an astronomical number of 'collisions'(13) before the biological time scale begins, producing highly overdamped systems, in which all motions are frictional (11, 12) driven by electrodynamics, more than anything else. Hardly any motions resemble the dreams of biologists (decades ago) in which uncharged atoms jump frictionless over barriers, as in ideal gases. These dreams are found in the versions of rate theory implicitly accepted even today in much of molecular and structural biology, but they are on their way to being replaced by realistic physics.

What could we expect if the quantitative success of the physical sciences could be replicated—even in crude approximation—in the biological sciences? Perhaps the qualitative 'arrow' descriptions of molecular mechanism found in nearly every textbook—that show impossibly smooth unidirectional trajectories of individual atoms—could be replaced by trajectories that show the reversals and complexities of thermal, nearly Brownian motion driven by electrodynamics, more than anything else. The smooth over-approximated trajectories could then be recognized as the averaged coarse grained results that they are. Quantitative analysis of the limitations of the averaging could begin, revealing the correlations that are and are not well described by each averaging procedure. The well-established methods of the theory of stochastic processes can be applied to the problems of nonequilibrium statistical physics on the atomic distance scale (say $> 10^{-11}$ meters) and the biological time scale (say $> 10^{-5}$ sec). For example, what are the statistical properties of the trajectories in the simulations of molecular dynamics? Are trajectories correlated? If so, how? In that way some of the qualitative ideas of atomic mechanism (found in so many papers in structural biology) could be replaced by quantitative predictions of biological function that are testable, transferrable—i.e., predictive with one set of

parameters—and satisfy the fundamental equations of electrodynamics, diffusion, and mass transport.

Of course, the reader might think that quantitative models are impossible in biology because of its complexity, and that reader may be right. But someone raised in the tradition of classical biophysics knows how the complexity of the nerve signal, from protein molecule to transmission of signals over meters of nerve, was unraveled into a set of differential equations. Nerve transmission was thought to be essentially chemical by a leader of British and world biophysics, a Nobel Laureate in Physiology (14)., It was reduced to equations and physics by a young research student doing clever controlled experimentation quantitatively with appropriate custom built equipment (15, 16) that eventually allowed molecular, even atomic description(17-19) recognized in a sequence of Nobel Prizes.

An essential embodiment of the classical tradition of biophysics has been the **_Biophysical Journal_** itself, launched so successfully with an important paper by Cole and Moore, Vol. 1, p. 1 (20). Anyone who reads the quantitative (almost frighteningly controlled) experimentation of the classical papers of physiology, then biophysics, (15, 16, 21-25) will be aware of the care with which each measurement or procedure was checked. Indeed, some of the papers are likely to overwhelm modern readers and exceed their patience because of the extensive detailed measurements needed to perform proper controls.

Applying physical methods to biology is limited as much by cultural restrictions in my view as by the complexity of biology. I do not believe that many biological systems are significantly more complex than the computers that run our smartphones or their integrated circuits. Classical biophysics shows that physical methods can produce nearly complete physical understanding of entire biological systems—almost without vitalism—on the atomic scale, e.g., the nerve signal linking sensory cells to spinal neurons, sometimes meters away. (The quantitative vagueness found in analysis of conformation change in proteins will no doubt be replaced by numbers in future years.)

One of the cultural restrictions of biology is the looseness of quantitative reasoning, the absence of quantitative controls found in the central contributions of much of structural and molecular biology, even molecular dynamics. This leads quickly, in my opinion, to detachment from the physical properties of trajectories of charged particles that have been so productively exploited by computational and semiconductor electronics. Let's hope the proper physics will prove as productive for biophysics and medicine as it has for electronics.

Davis, et al, (1) show in a most practical way that quantitative measurement—careful experimentation using custom built instrumentation—can have immediate implications for the biophysical measurements we do every day in the lab, promising significant advances in technique and understanding of biological function.

**Acknowledgement:** I am grateful for important suggestions from Brian Salzberg and Ardyth Eisenberg that significantly improved the manuscript.